\documentclass[rnote]{aa} 
\usepackage{graphicx}
\usepackage{epstopdf}

\usepackage{txfonts}
\begin{document}

   \title{Double-nucleus elliptical MCG-01-12-005 in an X-ray emitting cluster of  galaxies}

   \author{Roberto Nesci\inst{1}, Mariateresa Fiocchi\inst{1}, L. Bassani\inst{2}, 
   Pietro Parisi\inst{1}
          }

   \institute{INAF-IAPS, via Fosso del Cavaliere 100,
              I-00033 Roma, Italia\\
              \email{roberto.nesci@iaps.inaf.it}
         \and
            INAF-IASFBO
            via P. Gobetti 101, 40129 Bologna, Italia
             }

   \date{13 mar 2015}

 
\titlerunning {A double nucleus elliptical}
\authorrunning{R. Nesci et al.}

  \abstract
  { The scenario of galaxy formation is believed to follow a
structure that builds up from the bottom, with large galaxies being formed by several merging episodes of smaller ones. In this scenario a number of galaxies can be expected to be seen in the merging phase, with their external regions already mixed, while their  nuclei, with stronger self-gravitation, are still recognizable as such. 
   During a photometric monitoring of AGNs in the field of a long-exposure INTEGRAL pointing, we serendipitously found an elliptical galaxy in the center of the X-ray cluster (EXO 0422-086) with two nuclei. We performed surface photometry on our images and those of the SDSS archive and obtained slit spectra of both nuclei. 
Aperture photometry of the two stellar-like nuclei showed very similar colors in the SDSS image and in our Johnson BVRI images, which is typical of an elliptical galaxy nucleus. The spectra of the nuclei showed the typical absorption lines of an elliptical galaxy without appreciable emission lines. The redshifts derived from each nucleus were equal and fully consistent with the literature value (0.0397). We can  therefore exclude the possibility that one of the nuclei is a foreground star or a background AGN and consider this elliptical galaxy as a bona fide example of a galaxy merger.
}


   \maketitle
%

\section{Introduction}
The formation of elliptical galaxies is generally assumed to be the result of several merging events, as demonstrated by several simulations (see, e.g., van Wasselhove et al. 2012; Ascaso et al. 2014).
As the timescale of this process is estimated to be short, catching a galaxy during the capture of a satellite is very rare: the definition of a sample of merging galaxies is therefore useful to better constrain the details of the process from an observational point of view (e.g., Darg et al. 2010a, 2010b; Hwang \& Chang 2009).
We observed the galaxy  MCG-01-12-005 in the cluster EXO 0422-086 during an optical monitoring of a sample of blazars included in the field of view of a deep exposure (3 Ms) of a field in the Orion region with the INTEGRAL satellite.
The aim of our data-rights proposal on this field was to take advantage of the long integration time to better measure the hard X-ray flux of the blazars serendipitously included in the field and thus to improve the measure of their spectral energy distribution: to this purpose, we collected nearly simultaneous optical data with the Loiano 1.5m and Asiago 1.8m telescopes.  To select the targets, we used the RomaBZCat catalog (Massaro et al. 2009), where this source is indicated as a possible blazar of uncertain classification (BZU J0425-0833).

We found two point-like sources in the center of the galaxy, aligned in the N-S direction, separated by just 3 arcsec, with similar magnitudes: in the Sky Surveys plates (both from the US and the UK Schmidt telescopes) the nucleus of the galaxy seems elongated N-S, but the two sources are not resolved because the central part of this galaxy is saturated. 

The presence of a double nucleus was not reported in any previous optical study; the galaxy is listed in the catalog of interacting galaxies (as VV 760) by Vorontsov-Velyaminov (1977) mainly because of the nearby spiral galaxy MCG-01-12-006 (Bassani et al. 1999).
Our galaxy is the brightest member of a cluster, recognized as such only in 1993 as a result of its X-ray emission, which was detected with the EINSTEIN satellite (David et al 1993). Several subsequent papers were then devoted to studying the X-ray cluster properties, but few papers were dedicated to an optical study of its central galaxy. De Grandi et al. (1999) only reported a redshift, taken from an ESO Key Project. Bauer et al. (2000) performed a cross-correlation of ROSAT and NVSS catalogs, associating a radio flux of 114 mJy with the cluster. Belsole et al. (2005) performed a detailed spatial analysis of the cluster and of its central galaxy at arcsec resolution in X-rays with the {\it Chandra} satellite and in the radio (at 1.4, 4.9 and 8.4 GHz) with the VLA, finding a complex radio and X-ray morphology in the very center of  MCG-01-12-005, but used Palomar Survey plates for a comparison with the  optical morphology. Guzzo et al. (2009) reported the redshift of this galaxy in a comprehensive spectroscopic study of a sample of bright ROSAT X-ray clusters (REFLEX survey, Boeringer et al. (2004) ), but again with no mention of its double optical nucleus.

We  examined the images of this galaxy in the Sloan Digital Sky Survey (SDSS DR8 and later), which clearly showed the two nuclei, thanks to a spatial resolution and dynamic range much better than the DSS plates; but they were not identified as separate sources, and the catalog lists only a single entry. 
We therefore decided to perform a new photometric and spectroscopic analysis of the galaxy, obtaining slit spectra of both optical nuclei, to derive their redshift and understand their nature.

We note that MCG-01-12-005 is not included in the Galaxy Zoo merging galaxies catalog by Darg et al. (2010), which is based on the SDSS DR6 release, nor is it listed in the merging galaxies catalog by Hwang and Chan (2009).

Darg et al. (2010) reported a fraction of about 3\% for merging galaxies, with a higher fraction of spiral or elliptical galaxies than in the normal field. In their scheme, our galaxy should be best represented by the "approaching post-merger" class, where the progenitor cores are typically within 5 arcsec of each other in the images, quite comparable to the case of MGC-01-12-005. This class contains only 10\% of their merging galaxies sample, which are therefore the least common case.

Hwang and Chan (2009) used an automatic algorithm to measure the isophotal shapes instead of employing human inspection: they found that about 0.38\% of galaxies are mergers. Their classificaton scheme include mergers, close pairs, and close multiples, but  they do not have a double nucleus subset,  and their technique is not tailored to find them.
No detailed X-ray or radio study of the merging galaxies is presented in any of these papers.

As a final remark, we note that MCG-01-12-005 resembles a cD galaxy,
that is, a giant elliptical galaxy located near the center of a cluster of
galaxies: cD galaxies are believed to have grown so large through galaxy mergers that they spiral in to the center of the cluster. The merger
hypothesis  is supported by frequent observations of  complex  nuclei
structures in cD galaxies: for example,  Hoessel (1980) analyzed the
central galaxies of 108 nearby (z$<$0.1) Abell galaxies and found that 33
contain double or multiple nuclei. It is therefore not surprising that 
MCG-01-12-005 also shows a double nucleus.
This source is interesting for the wealth of optical radio and
X-ray data that are available despite the confusion in the literature
about the entire system (cluster misclassified as a blazar, MCG -01-12-005
misclassified as an AGN).

\section{Observations}
We observed the galaxy with the Loiano 1.5m telescope equipped with the Bologna Faint Object and Spectrograph Camera (BFOSC), using grism n.4 (dispersion 4.0 A/pixel, 2" slit) , and the Asiago 1.8m equipped with the Asiago Faint Object and Spectrograph Camera (AFOSC) grism n.4 (dispersion 5.0 A/pixel, 1.69" slit). The slit was aligned with the two nuclei to cover both spectra in a single shot. 
Direct images were also obtained at Asiago with AFOSC in the BVRI Johnsons-Cousins filters, with a scale of 0.57"/pixel, and  at Loiano  with BFOSC with a quite similar scale of 0.58"/pixel:
all observations were performed in service mode.

The observations  log is collected in Table 1. 

\begin{table}
\caption{Observations log}             
\label{table:1}      
\centering                          
\begin{tabular}{c c c }        
\hline\hline                 
date & Telescope & filters/grism  \\    
\hline                        
   2013-01-31& Asiago & $BVRI$  \\      
   2013-09-24& Loiano & $BVRI$   \\
   2013-12-09 & Loiano & grism-4, 2" \\
   2014-10-30 & Loiano & grism-4, 2" \\
   2014-10-30 & Asiago & grism-4, 1.69" \\
   2014-11-02 & Asiago & grism-4, 1.69" \\
\hline                                   
\end{tabular}
\end{table}

\subsection{Astrometry}
The angular resolution of the SDSS images is appreciably better than that of our Asiago or Loiano images: we therefore used the best-exposed SDSS image (r filter) to derive the astrometry of the nuclei. The southern spot is at 04h 25m 51.33s -08 33' 39.9", the northern spot at 51.31s 37.1". The astrometric solution for our Asiago and Loiano images using the GSC2.3.2 catalog provided coordinates equal to those of the SDSS.

The angular separation of 3" at the galaxy distance of 160 Mpc (from standard cosmology) gives a distance between these structures of 2.4 kpc, if located on the plane of the sky.

We compared these positions with those of the two radio spots detected at 4.9 GHz with the VLA reported in Belsole et al. (2005);  the last digits for the coordinates are SW 51.238s 37.51", NE 51.325s 35.96". The radio spots are therefore aligned at about 45 degrees from the N-S direction, separated by only 2": the optical northern nucleus is nearly  between the radio spots, while the optical southern nucleus is definitely located away (south) from them. 

We remark that a systematic shift of just 1" of the VLA radio image with respect to the optical image would be enough to put the optical source between the two radio spots.
Unfortunately, there are no other strong X-ray and radio sources in the field to check for possible systematic offsets between the coordinates of the optical, X-ray, and radio sources at arcsec accuracy. If the double radio source is related to the central black hole in one of the cores, then most likely it must be in the northern optical spot. We show in Fig. 1 the radio and X-ray map of the galaxy center taken from Fig. 2 by Belsole et al. (2005) with the positions of the optical nuclei overplotted. 

We then tried to estimate the position of the galaxy center from the isophotal lines of the galaxy: an isophotal map, built with SAOImage from the {\it r} image of the SDSS,  is given in Fig. 2. To determine the light barycenter, we excluded the inner isophote, which is dominated by the two bright point sources, and used the subsequent five isophotes: The centers of the second, third, and fourth isophote are definitely on the northern spot, while the centers of the sixth and seventh (more external) isophotes are located between the two spots. 

Overall, we conclude that the light barycenter of the galaxy is more consistent with the northern spot, which also has a slightly more diffuse appearence (FWHM of 6 instead of 4 pixels), as expected for the core of an elliptical galaxy. 

If the galaxy is not at rest with respect to the intracluster medium (ICM), as found for the central galaxy in a number of clusters by Coziol et al. (2009), and if the radio source is anchored to the ICM, then the shift between the optical nucleus and radio source might have been produced in just a few million years.

   \begin{figure}
   \centering
   \includegraphics[width=8cm]{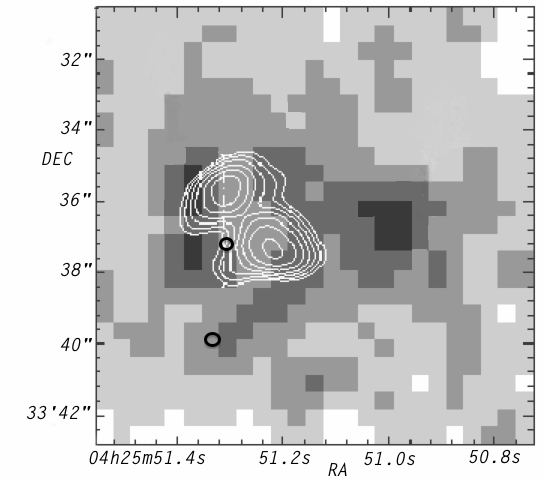}
      \caption{Radio, X-ray, and optical map of the galaxy center. Black circles are the optical nuclei, gray shades the Chandra X-ray emission, white lines the radio isophotes.  
              }
         \label{radio}
   \end{figure}
%

%
   \begin{figure}
   \centering
   \includegraphics[width=9cm]{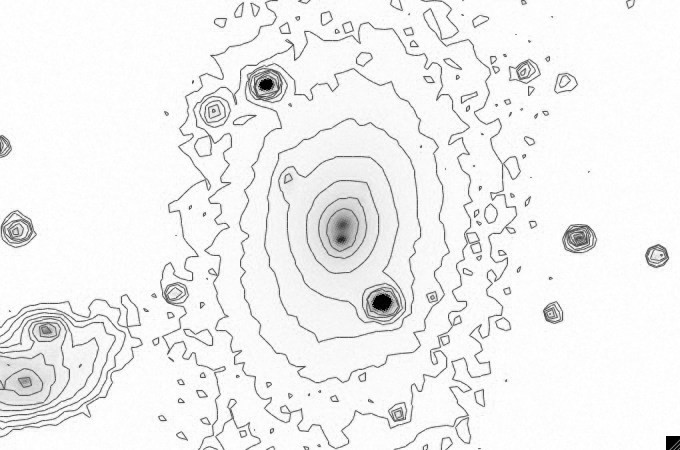}
      \caption{Isophotes of MCG-01-12-005 from the r image of the SDSS. Levels are at 0.030
0.052, 0.090, 0.155, 0.269, 0.467, 0.808, and  1.400 flux units.
              }
         \label{isofote}
   \end{figure}
%
\subsection{Spectroscopy}
We extracted the spectra of the two nuclear sources with IRAF/apall, using a slit length of 3.4" ($\pm$2 pixels), which is comparable to the FWHM. Wavelength calibration was made with an Hg-Cd lamp (Asiago) or He-Ar (Loiano), as is routinely used in these observatories. The Asiago and Loiano spectra gave very similar results,  making us confident of their physical interpretation. 

The two nuclei only showed absorption lines, as is typical of the stellar population of an elliptical galaxy, as expected. No emission lines were detected in either spectrum. A redshift of 0.040 was derived using the CaII H-K , G-band, Mg-b triplet,
and the NaI D doublet (blended), in full agreement with the results by Guzzo et al. (2009) and Huchra et al. (2012).
The portion of the Asiago spectra including these lines is reported in Fig. 2, with the main absorption lines flagged.

   \begin{figure}
   \centering
   \includegraphics[width=9cm]{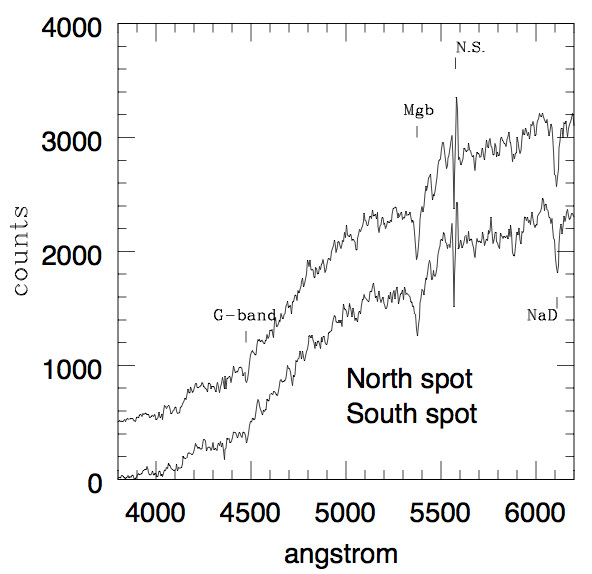}
      \caption{Spectra of the southern (below) and northern (above) spots with the
      main absorption lines, both giving a redshift of 0.04.
              }
         \label{spettri}
   \end{figure}

It is therefore evident that neither of the two nuclei is a background QSO/BLLac or a foreground star: MCG-01-12-005 looks like a bona fide giant elliptical galaxy with two nuclei.

\subsection{Photometry}

The public {\it ugriz} images of the SDSS DR8 were obtained on 2007-02-07 with a scale of 0.40"/pixel. In this database, as well as in the following DR9 and DR10, the photometry of the galaxy is flagged as unreliable, and the two nuclear sources are not reported as separate sources. 
Therefore we measured the SDSS magnitudes of the two sources with iraf/apphot, using as  photometric radius for all filters the average FWHM of the stellar images; the sky value was taken from a corona with an inner radius 45" and width 10", chosen to be substantially outside the galaxy boundary as evaluated from the visible image of the galaxy in the r band. The mode of the histogram of the sky values was assumed as  the sky level in all cases. 
The same criteria were used to  perform the photometry in the  B, R, and I Loiano images; in the Asiago images the seeing was definitley poorer and the two sources were more blended, so we did not reduce their photometry. The results are collected in Table 2: Col. 1 is the filter, Col. 2 the southern nucleus, Col. 3 the northern nucleus, Col. 4 the difference, and Col. 5 the error, given by the rms deviation of the comparison stars from the calibration linear fit. 

Detecting possible magnitude variations of each nucleus between the two epochs (2007 and 2013) is challenging, both because of the conversion between Sloan and Johnson-Cousins filters and because of the different seeing conditions, which include a variable portion of the surrounding galaxy in the point source. 
However, given the lack of emission lines in their spectra, we do not expect that these nuclei are variable.

\begin{table}
\caption{SDSS and Loiano aperture photometry}             
\label{table:2}      
\centering                          
\begin{tabular}{ c c c c c}        
\hline\hline                 
Filter & South& North &diff  &err \\    
\hline                        
   u& 19.97 & 20.42& 0.45 &0.10  \\  
   g& 18.08 & 18.18 & 0.10 & 0.06\\      
   r& 17.03& 17.11  & 0.08 & 0.05\\
   i& 16.57& 16.64 & 0.07 & 0.05 \\
   z& 16.16& 16.21 & 0.05 & 0.07\\
   B& 18.31 & 18.36 & 0.05 & 0.07\\
   R& 16.83 & 16.89 & 0.06 & 0.04\\
   I& 15.97 & 16.03 & 0.06 & 0.05\\
   
\hline                                   
\end{tabular}
\end{table}

The most remarkable photometric result from the SDSS images is that the southern spot is much brighter in the {\it u} band than the northern spot, while at longer wavelengths the two sources have rather similar colors. This is confirmed by the BRI magnitudes of the Loiano images.
As we noted above,  the radio source seems associated with the northern spot, suggesting that it might be an active galactic nucleus and therefore likely to have an UV excess, while the observation gives an opposite result, which is rather puzzling. A possible explanation is that the radio emission is just a remnant of a past nuclear activity.  However, the u-r color of both nuclei is $\sim$ 3, while it is generally lower than 1.4 in bona fide Blazars (see Massaro et al. 2012).

The galaxy was barely detected by GALEX in the NUV band (2270 \AA) at mag 19.49$\pm$0.14, but its angular resolution (5 arcsec) is insufficient to resolve the two sources. In the GALEX image, the nearby MCG-01-12-006 Seyfert galaxy is much brighter in the UV   (NUV=17.6, FUV=17.9).

\section{Conclusions}
The galaxy MCG-01-12-005 is the giant elliptical in the center of the X-ray cluster EXO 0422-086. A detailed X-ray study of the cluster with Chandra (Belsole et al. 2005) ruled out the presence of an X-ray emitting AGN, showing that the emission is due to the intracluster medium and the interstellar medium of the central elliptical galaxy.
Our optical observations showed two clearly separated nuclei 3" that are located apart and are nearly aligned in the N-S direction. The optical spectra of these nuclei are very similar, with no emission lines, showing the absorption features typical of an elliptical galaxy: both sources have the same redshift, z=0.04, consistent with previous  data in the literature, which did not mention these two nuclei.  Images from the SDSS confirm the presence of the two nuclei. 
A morphological study of the isophotes shows that  the northern nucleus is located in the light barycenter of the galaxy, suggesting that it is also in the mass center. A radio high-resolution image taken with the VLA (Belsole et al. 2005) shows two small sources separated by few arcsecs, nearly symmetric to the northern optical nucleus.

The southern nucleus is most likely the remnant of a satellite galaxy with a velocity vector nearly in the plane of the sky. In principle, it could  also be interpreted as an independent galaxy of the cluster that by chance is located in front of the giant elliptical: the isophotes do not show any trace of the external part of this putative galaxy, however, so we prefer the interpretation of a  merger in the final phase, when the outskirts of the smaller galaxy have already been absorbed by the giant, while the core still retains its shape because
of its stronger self-gravity.

Finally, our spectroscopic data confirm the lack of emission lines in the galaxy, as already found by Guzzo et al. (2009). The (u-r) color for both nuclei is definitely higher than the typical value for an AGN.  We therefore conclude that there is no AGN currently active in MCG-01-12-005. The faint double radio source might be the remnant of a past activity, now left off center by the peculiar motion of the galaxy in the intracluster medium.

\begin{acknowledgements}
We thank the Asiago and Loiano Observatories Time Allocation Committes for the allocated telescope times; Roberto Gualandi (Loiano) and Paolo Ochner (Asiago) for making the observations in service mode. We thank the anonymous referee for useful suggestions.

\end{acknowledgements}

\end{document}